# The unexpectedly large dust and gas content of quiescent galaxies at z > 1.4

R. Gobat[1], E. Daddi[2], G. Magdis[3,4], F. Bournaud[2], M. Sargent[5], M. Martig[6], S. Jin[2,7], A. Finoguenov[8,9], M. Béthermin[10], H.S. Hwang[1], A. Renzini[11], G.W. Wilson[12], I. Aretxaga[13], M. Yun[12], V. Strazzullo[14], F. Valentino[3]

**Early type galaxies (ETG) contain most of the stars present in the local Universe and, above a stellar mass of $\sim 5 \times 10^{10}$ M$_\odot$, vastly outnumber spiral galaxies like the Milky Way. These massive spheroidal galaxies have, in the present day, very little gas or dust in proportion to their mass[1], and their stellar populations have been evolving passively for over 10 billion years. The physical mechanisms that led to the termination of star formation in these galaxies and depletion of their interstellar medium remain largely conjectural. In particular, there are currently no direct measurements of the amount of residual gas that might be still present in newly quiescent spheroids at high redshift[2]. Here we show that quiescent ETGs at $z \sim 1.8$, close to their epoch of quenching, contained at least 2 orders of magnitude more dust at fixed stellar mass than local ETGs. This implies the presence of substantial amounts of gas (5–10%), which was however consumed less efficiently than in more active galaxies, probably due to their spheroidal morphology, and consistently with our simulations. This lower star formation efficiency, and an extended hot gas halo possibly maintained by persistent feedback from an active galactic nucleus (AGN), combine to keep ETGs mostly passive throughout cosmic time.**

The presence of quiescent galaxies, with very low relative star formation rates (SFR), has been established up to $z \sim 3$[3,4]. Their existence a mere 2 Gyr after the Big Bang implies that, in at least some regions of the Universe, the processes responsible for the cessation of star formation were already very efficient. The termination of star formation in ETGs is usually attributed to the removal of cool gas reservoirs (e.g., by stellar or quasar feedback[5]) and/or by the suppression of gas infall and cooling (e.g., by virial shocks or AGN feedback[6,7]). Alternatively, the growth of bulges and stellar spheroids is thought to stabilise gas reservoirs, making star formation inefficient compared to disk galaxies[8,9]. If the latter plays an important role in galaxy quenching, we might expect substantial reservoirs of untapped gas to exist in galaxies that have recently turned quiescent. Detecting this residual gas at high redshift, close to the epoch of quenching for massive quiescent galaxies, is however very challenging[2] and all attempts have so far been unsuccessful.

Deep survey fields allow us to circumnavigate this issue. By combining data from large numbers of undetected quiescent galaxies, their average far-infrared (FIR) emission can be observed, which traces the cold dust present in their interstellar medium (ISM) and, indirectly, their SFR. We select 977 isolated, high-mass ($\langle M_\star \rangle = 1.1 \times 10^{11}$ M$_\odot$) quiescent galaxies in the 2 deg$^2$ COSMOS field by combining two different photometric criteria (*BzK*[10] and *UVJ*[11]) and keeping only objects that are individually undetected at observed mid-infrared (MIR) wavelengths (Methods). These criteria ensure that the sample contains only the least star-forming galaxies at z = 1.4 – 2.5, with clear early-type morphologies (as implied by a high median Sérsic index n $\sim$ 3.5; Methods). We extract cutouts centred at the position of each galaxy from the 24$\mu$m (MIR), 100 – 500$\mu$m (FIR), 0.85 – 1.1 mm (sub-millimetre; sub-mm), 10 and 20 cm (radio) observations of COSMOS and perform a median stacking analysis at each wavelength. After correcting for the contribution from satellite galaxies and unassociated neighbours in the line of sight (Methods), we obtain a clean measure of the FIR and radio emission of the quiescent galaxies, with detections in all but two bands (100 and 160$\mu$m) ranging in significance from 3.5$\sigma$ to 11.9$\sigma$ (Figure 1).

We model both the FIR emission from cold dust[12] and non-thermal radio emission from star formation, based on the FIR flux (Methods). While the models underpredict the observed radio fluxes by a factor $\sim$3 (corresponding to an excess of $\sim 5 \times 10^{22}$ W/Hz), they reproduce the FIR SED well, implying a luminosity of $L_{IR} = 2.9 \pm 0.9 \times 10^{10}$ L$_\odot$. Furthermore, sampling the peak ($\sim$150 $\mu$m rest-frame) and Rayleigh-Jeans tail ($> 400\mu$m rest-frame; Figure 1) of the FIR emission allows for an accurate determination of the dust temperature, $T_{dust} \sim 21 - 25$ K, and mass, $M_{dust} = 1^{+0.6}_{-0.4} \times 10^8$ M$_\odot$. The temperature of the cold dust component, as suggested by the lack of detection at $\sim 100 - 160$ $\mu$m but not at $\geq 250$ $\mu$m, is similarly up to 10 K lower than the main sequence of star formation (MS) at the same $L_{IR}$[22], among the lowest values for star forming (SF) galaxies at any redshift but consistent with that of local ETGs[23]. This low temperature also confirms that the measured

[1]School of Physics, Korea Institute for Advanced Study, Hoegiro 85, Dongdaemun-gu, Seoul 02455, Republic of Korea
[2]CEA, IRFU, DAp, AIM, Université Paris-Saclay, Université Paris Diderot, Sorbonne Paris Cité, CNRS, F-91191 Gif-sur-Yvette, France
[3]Dark Cosmology Centre, Niels Bohr Institute, University of Copenhagen, Juliane Mariesvej 30, DK-2100 Copenhagen, Denmark
[4]Institute for Astronomy, Astrophysics, Space Applications and Remote Sensing, National Observatory of Athens, GR-15236 Athens, Greece
[5]Astronomy Centre, Department of Physics and Astronomy, University of Sussex, Brighton, BN1 9QH, UK
[6]Astrophysics Research Institute, Liverpool John Moores University, 146 Brownlow Hill, Liverpool L3 5RF, UK
[7]School of Astronomy and Space Science, Nanjing University, Nanjing 210093, China
[8]Max-Planck-Institute for Extraterrestrial Physics, Giessenbachstrasse, D-85748 Garching, Germany
[9]Department of Physics, University of Helsinki, PO Box 64, FI-00014 Helsinki, Finland
[10]Aix Marseille Univ., CNRS, LAM (Laboratoire d'Astrophysique de Marseille) UMR 7326, F-13388, Marseille, France
[11]INAF-Osservatorio Astronomico di Padova, Vicolo dell'Osservatorio 5, I-35122 Padova, Italy
[12]Department of Astronomy, University of Massachusetts, Amherst, MA 01003, USA
[13]Instituto Nacional de Astrofísica, Óptica y Electrónica (INAOE), Aptdo. Postal 51 y 216, 72000 Puebla, Pue., Mexico
[14]Department of Physics, Ludwig-Maximilians-Universität, Scheinerstr. 1, D-81679 München, Germany

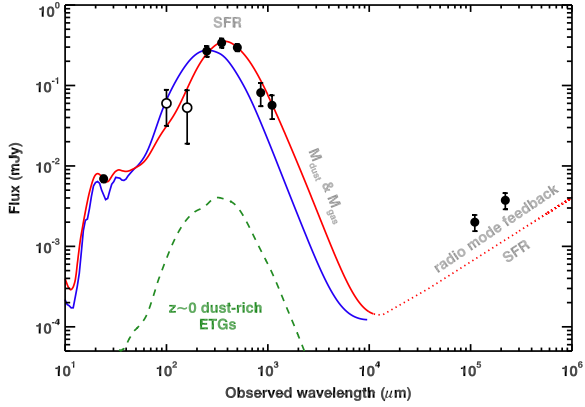 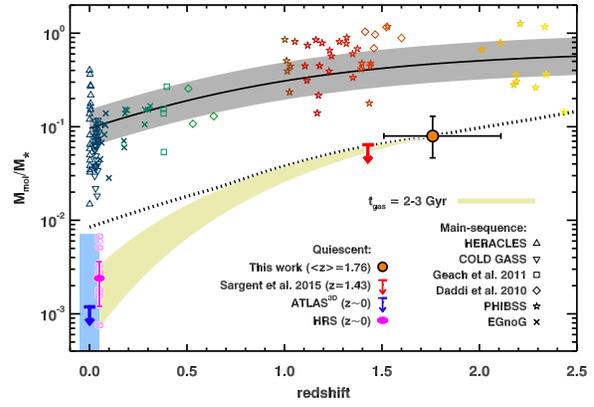

Figure 1: Mid-infrared to radio SED of 24 $\mu$m-undetected pBzK+UVJp galaxies, from stacked cutout images (filled circles, the empty circles representing <$3\sigma$ non-detections), with best-fit model from Draine & Li[12] (solid red curve, dotted when extended into the radio regime; Methods). For comparison, the "best-fit" standard MS template[13], which has hotter dust, is shown in blue, and the median best-fit dust component to the FIR SED of local *dust-rich* ETGs as a dashed green curve. The latter model has been redshifted to z = 1.76 and rescaled to the same stellar mass. All models have been broadened using the redshift distribution of the sample as kernel. The error bars on the MIR, FIR, and radio fluxes show the bootstrap uncertainties derived from the stack.

Figure 2: Evolution of the molecular gas fraction $M_{mol}/M_\star$ as a function of redshift for both quiescent and MS galaxies. The latter include low-redshift[14,15], intermediate-redshift[16,17], and z > 1 MS galaxies[18,19,13]. The distant ETG sample is shown by an orange dot, with error bars from the SED modelling (Methods), and the upper limit constraints from CO(2-1) observations of a massive ETG at a similar redshift[2] by a red arrow. The blue arrow marks the median upper limit on this ratio for ETGs in the ATLAS$^{3D}$ survey, with the light blue shaded region showing the range of values for the detected sub-sample (22%) of ATLAS$^{3D}$. For comparison, local ETGs in the *Herschel* Reference Survey (HRS) are shown by magenta ellipses: open for gas masses from CO measurements[20], including upper limits, and filled for dust masses[1] converted using the same G/D, with the error bar showing the median absolute deviation of the sample. The black curve and grey strip show the evolution of an average MS galaxy[21] with stellar mass $5 \times 10^{10}$ M$_\odot$, and a 0.2 dex scatter approximately corresponding to the 1$\sigma$ scatter of the MS. The dotted line shows the relation[21] offset by a factor $\sim$ 6 and for a stellar mass of $\log M^\star(z) + \Delta M$, where $\Delta M$ is the offset between the median mass of the total pBzk+UVJp sample and the $M^\star$ of the stellar mass function of quiescent galaxies at the median redshift of the sample. Finally, the green-yellow region shows gas consumption in the closed box case (no gas inflow or outflow), with a timescale $t_{gas}$ between 2 and 3 Gyr (lower and upper edge, respectively).

FIR emission does not originate from star forming satellites, which would have $T_{dust}$ values closer to the MS ($T_{dust,MS} \sim$ 30 K; Methods). These distant ETGs appear however to be much more dust rich than their local counterparts, with $(M_{dust}/M_\star)_{z=1.76} \sim 8 \times 10^{-4}$ compared to $(M_{dust}/M_\star)_{z=0} \sim 10^{-6} - 10^{-5}$ locally[23,1], a difference of $\gtrsim 2$ orders of magnitude. We convert the measured dust mass into a gas mass assuming a gas-phase metallicity close to the solar value and a metallicity-dependent gas-to-dust ratio (G/D)[13], which yields $M_{gas} = 9.5^{+5.6}_{-3.5} \times 10^9$ M$_\odot$ (Methods). In nearby quiescent galaxies for which such measures exist, the relative content of molecular ($H_2$) to atomic ($H_I$) hydrogen gas ranges from $M(H_2)/M(H_I) \sim$ 0.1 to 100, with molecular gas nevertheless dominating by a factor $\sim$5 on average[24,25]. We thus assume that distant quiescent galaxies are similar to their local counterparts in this regard (Methods) and that $M_{gas}$ hereafter refers to molecular hydrogen rather than the combined neutral and $H_2$ gas mass.

The large amount of dust we detect, and the inferred gas mass, implies that high redshift ETGs are remarkably gas rich, with up to 11% of their baryonic mass being in the form of gas. As with the dust fraction, this represents an increase of more than two orders of magnitude with respect to the local Universe[1] (Figure 2), even when accounting for differences in stellar mass between local and high-redshift samples, and a cosmic evolution that vastly exceeds the similar increase in the gas fraction of SF galaxies. In "normal" SF galaxies, the FIR and sub-mm emission is a relatively pure tracer of star formation. If the observed cold dust emission does indeed arise from star formation activity, converting[26] $L_{IR}$ into a SFR then yields $4.8^{+1.8}_{-1.3}$ M$_\odot$ yr$^{-1}$, in excellent agreement with the SFR of z $\sim$ 1.5 ETGs derived from optical spectroscopic diagnostics[27], which probe different timescales (a few tens of Myr for spectroscopic line emission, several times that for FIR emission). Given the large stellar mass of these galaxies, it implies that the distant ETGs have a sSFR that is a factor $\sim$ 30 lower than that of active galaxies on the MS at the same epoch (Figure 3). The presence of both such large gas reservoirs and low SFR would imply that the star formation efficiency (SFE) of these quiescent galaxies is 2 – 3 times lower than that of disk galaxies and comparable to that of local ETGs[28] (Figure 3). This value is conservative and thus could be even lower, if part of the FIR emission of these distant ETGs arises from, e.g., heating of cirrus dust by an evolved stellar population consistent with the median ETG age at this redshift. Whether massive, main sequence stars or post-main-sequence stars (i.e., ongoing of past star formation, respectively) are the dominant heating source of the dust in these ETGs then depends on its spatial distribution (concentrated in and around molecular clouds or diffuse), which is still unknown. The good agreement between optical and IR-based SFRs[27], though, suggests that residual star formation is an important contributing mechanism to the FIR emission. A low SFE is also consistent with the low dust temperature and small value of the dust-mass weighted luminosity $\langle U \rangle$ estimated from the IR SED, since the intensity of the radiation field can be related to the SFE and metallicity by the relation $\langle U \rangle \propto SFE/Z$[13]. It also implies a longer gas consumption timescale, $t_{gas} = 2-3$ Gyr compared to $\lesssim 1$ Gyr for MS galaxies. A low efficiency in ETGs is predicted by high-resolution hydrodynamical simulations[8,29] (Methods), in which the SFE of a forming ETG, which starts off as a high-redshift disk galaxy, decreases by a factor 5 – 10 once the gas drops below a critical value (Figure 4). In our simulations, of gas disks embedded in spheroids, this corresponds to a gas fraction of 20 – 25%. This effect is linked to the presence of the spheroidal stellar component which, unlike a stellar disk, does not destabilise the gas disk. Conversely, in galaxies containing more than $\sim 25\%$ of gas, regardless of their stellar structure, the gas remains unstable and capable of forming stars efficiently.

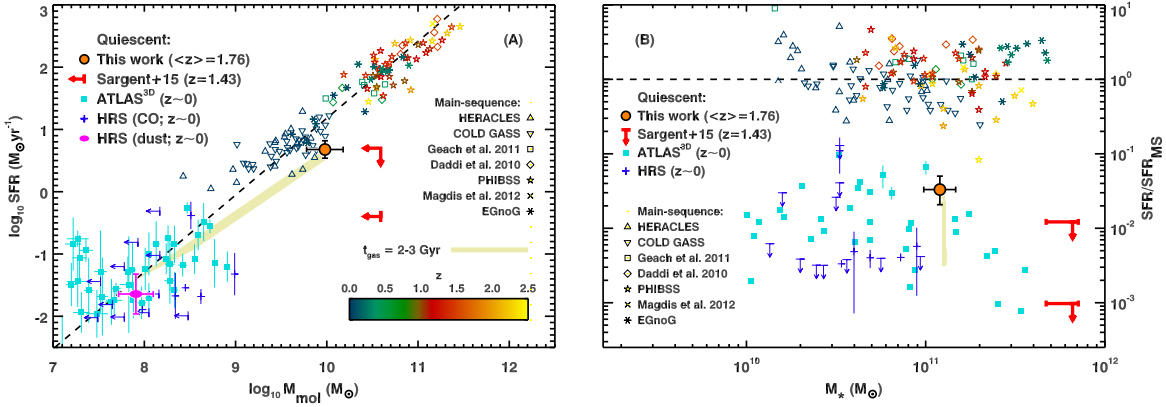

Figure 3: *A):* correlation between the molecular gas content and SFR of galaxies, including the same MS samples as in Figure 2. Our median measurement, shown as a orange filled circle with error bars (Methods), is offset by $\sim 3\sigma$ from the average SFR – $M_{mol}$ relation for MS galaxies[21] (black dashed line). For comparison, the upper limit constraints from CO(2-1) observations of a massive ETG at a similar redshift[2] is shown by red arrows, local ETGs from the HRS[20,1] as blue crosses and arrows (in the case of upper limits on $M_{mol}$), and independent measurements from the local ATLAS$^{3D}$ sample[24,31], when more than one dex below the MS in SFR, as light blue squares. As in Figure 2, the average SFR and dust-derived gas mass of HRS ETGs is indicated by a filled magenta ellipse, with error bars showing the median absolute deviation of the sample. The green-yellow region shows the same closed-box evolution from z = 1.76 to z = 0, as in Figure 2. *B):* SFR of star forming and quiescent galaxies as a fraction of the MS one[21], for the same data as in *A)*.

Our observations suggest that this process does indeed occur in high-redshift ETGs once the gas fraction has been reduced enough, and that the suppression of star formation thus does not require the complete removal of cool gas reservoirs. In principle, a passive evolution with no further accretion and a low SFE could then account for the low gas fractions of local ETGs, compared with their high-redshift counterparts, while only increasing their stellar mass by a small fraction between $z \sim 1.76$ and z = 0 (Figure 2). However, gas in local ETGs is likely to have had a more complex history. Indeed, several channels exist through which these galaxies could have been replenished with gas. Given ETGs ages of < 3 Gyr inferred from spectroscopy[27], we can estimate that $\sim 4.5 \times 10^{10}$ M$_\odot$ of metal-enriched gas were already returned to the ISM by stellar evolution, at a rate of $\lesssim 2.9$ M$_\odot$ yr$^{-1}$ at z = 1.76, with another $\sim 10^{10}$ M$_\odot$ produced over the next 10 Gyr assuming passive evolution (Methods). Gas is also expected to be reaccreted through mergers with gas-rich satellite galaxies and through the cosmic web. This suggests that there must exist other processes to keep the cold gas fraction low in these ETGs for the rest of their cosmic evolution, either by efficiently removing the surplus gas or preventing it from cooling. There are three likely such "maintenance" mechanisms: first, the host halos of these distant ETGs are, at $\gtrsim 10^{13}$ M$_\odot$, already massive enough to support a hot atmosphere, as evidenced by X-ray stacking (Methods). This hot gas can efficiently prevent the infall of cold gas onto the central galaxy[6]. The X-rays emitted by the halo can also remove, on short timescales and through sputtering, the dust produced by stellar evolution, thus contributing to the evolution of the dust content of ETGs between z > 1.4 and z = 0 (Methods).

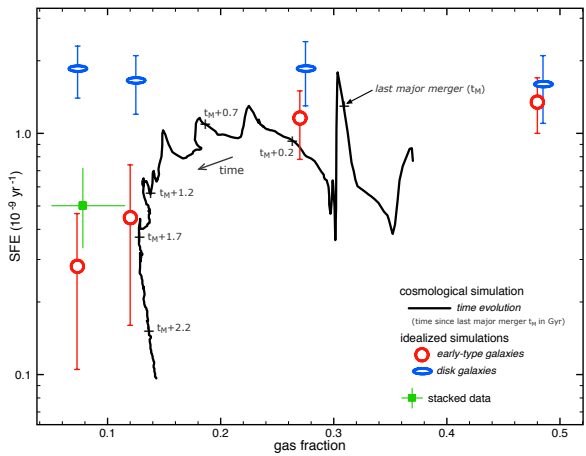

Figure 4: SFE as a function of gas fraction in high-resolution hydrodynamic simulations. The black line shows the evolutionary track of an ETG with an embedded gas disk in a cosmological zoom-in simulation, with initial peaks of SFE corresponding to galaxy mergers, followed by gas consumption and a strong decrease of the SFE once the gas fraction gets lower than 20-25%. This effect, which has been dubbed "morphological quenching", does not occur in galaxies whose stellar structure remains disk-like[8]. The galaxy already has an ETG morphology at the beginning of the track (Methods) but only quenches ~ 2 Gyr later when enough gas has been consumed. The red and blue symbols are from idealised simulations of galaxies of ETG and disk-dominated galaxies, respectively, where both gas structure and star formation are resolved at the parsec scale[32] (Methods). The error bars correspond to the r.m.s. fluctuation of the SFE between various snapshots of a given simulation. These confirm that the SFE drops below gas fractions of 20-25% only for ETGs, at a level consistent with our observations (filled green square with error bars). Local ETGs from ATLAS[3D] and HRS have SFEs similar to the idealised simulations and high redshift sample, but with much lower gas fractions ($3 \times 10^{-3}$ on average).

Second, the "radio-mode" feedback from AGN-driven jets (as tantalisingly suggested by the systematic radio excess, which would imply a duty cycle of nearly 1, as opposed to 1% locally; Figure 1 and Methods) can not only heat the halo gas but also regularly drive away a fraction of the cold gas reservoir[7]. Finally, gravitational interaction between satellites and the halo can also provide a source of heating for the gas[30].

The large amounts of dust we detect in our $z > 1.4$ sample, through its FIR emission, implies similarly large quantities of molecular hydrogen. This $H_2$ gas would thus appear to be abundant enough in these galaxies to be detectable with the Atacama Large Millimeter Array (ALMA) through sub-mm line tracers, although the confirmation of even a single galaxy can be expected to require several hours of observation. Nevertheless, our result indicates that semi-direct studies of gas in high-redshift quiescent galaxies are already feasible, although the large time expenditure required and still-unknown dependencies of the dust and gas fractions on stellar mass or environment could make their design challenging.

# Acknowledgements

The authors thank S. Lianou for providing models of dust emission in local ETGs and V. Smolčić for the 3 GHz radio data. S.J. acknowledges China Scholarship Council funding. The new simulations presented in this work were performed on GENCI resources (allocations 2016-04-2019 and 2017-04-2192).

# Author Contributions

R.G. and E.D. devised the project. R.G. analysed the data and wrote the manuscript. G.M. modelled the FIR emission. F.B. and M.M. carried out and analysed the simulations. M.S. and M.B. provided some of the theoretical framework. S.J. provided the MIR catalogue. A.F. analysed the X-ray observations. G.W.W., I.A., and M.Y. provided sub-mm data. H.S.H., A.R., V.S., and F.V. provided critical feedback that helped shape the manuscript.

# Correspondence

Correspondence and requests for materials should be addressed to R. Gobat (`rgobat@kias.re.kr`).

# Methods

## 1 Conventions

Stellar masses and star formation rates (SFRs) derived from photometry assume a Salpeter[33] stellar initial mass function, in accordance with the literature on IMF variations in ETGs[34,35,36]. We assume for consistency the same IMF for galaxies on the main sequence of star formation (MS). For distances, cosmic times, and cosmological simulations, we assume a ΛCDM cosmology with $H_0 = 70$ km s$^{-1}$ Mpc$^{-1}$, $\Omega_M = 0.3$, and $\Omega_\Lambda = 0.7$[37,38].

## 2 Sample selection

We select galaxies from the COSMOS field[39] using public UV-to-near infrared (NIR) photometry[40,41] and a new point-source-matched catalogue (Jin et al., in prep.; 8 $\mu$Jy r.m.s.) based on public *Spitzer*/MIPS 24 $\mu$m data[42]. We use spectroscopic redshifts from zCOSMOS[43] when available and photometric redshifts[41,44] otherwise. Quiescent galaxies at $1.4 < z < 2.5$ are first selected using the *BzK* criterion[10], using photometry from the original 2010 catalogue[40], for which colour selections have been extensively tested. The rest of the analysis makes use of more recent UltraVISTA photometry[41], however. In this first step, galaxies undetected at $<3\sigma$ in the B band, implying a B – z colour compatible with the locus of quiescent galaxies, are also considered. Photometric redshifts calibrated for passive stellar populations[44] are then used to compute rest-frame U-, V- and J-band magnitudes for the initial sample, so as to discard from the sample galaxies not also selected as quiescent by the high-redshift *UVJ* criterion[45]. As illustrated in Supplementary Figure 1, this double selection minimises the number of interlopers due to photometric uncertainties and highly-reddened star-forming galaxies. We then discard from the sample sources that are detected at $\geq 3\sigma$ significance in the 24 $\mu$m catalogue, about 14% of the sample, as it implies either considerable star formation (SF) or AGN activity. We then keep only galaxies with stellar masses higher than $\log M_\star = 10.8$ M$_\odot$, within uncertainties, so as to ensure that the stellar mass distribution of this sample is similar to that of a spectroscopically observed subsample[27] (Supplementary Figure 2) for which accurate redshifts and stellar population parameters (age, SFR, metallicity) were estimated. This mass limit also corresponds to twice the completeness limit of the catalogue for $z > 1.5$ quiescent galaxies[46]. We also ignore possible galaxy pairs, i.e., with concordant redshifts (within $\Delta z_{\text{phot}} \leq 0.2$) and separated by less than 60" ($\lesssim$ 30% of the sample), to simplify the modelling of the far-infrared (FIR) emission (Supplementary Section 3). The final sample contains 977 passive galaxies selected from a combined BzK+UVJ diagram, with a median redshift of $\langle z \rangle = 1.76$ and a median stellar mass of $\langle M_\star \rangle = 1.1 \times 10^{11}$ M$_\odot$.

## 3 Far-infrared stack

The COSMOS survey includes imaging data at 24 $\mu$m[42] (*Spitzer*/MIPS), 100 and 160 $\mu$m (*Herschel*/PACS, from the PEP survey[47]), 250, 350 and 500 $\mu$m (*Herschel*/SPIRE, from the HerMES survey[48]), 850 $\mu$m[49] (JCMT/SCUBA2), 1.1 mm[50] (ASTE/AzTEC), 3 GHz[52] and 1.4 GHz[51] (VLA). From these data we extract cutouts centred on each source in the sample with sizes 5 times larger than the beam FWHM. These are then combined to create a median 2D image in each band (Supplementary Figure 3), using a thousand bootstrap resamplings of the data with sizes of half the initial sample to estimate the variance of the sample. At each bootstrap iteration, we estimate the flux of the central ETG by assuming that the resulting median image is produced by a combination of point-like emission, contribution from star-forming satellites, and a background term. The FIR emission from the ETGs' satellite halos is estimated in the following way[46]: we select, within 120" of each ETG (hereafter, "central"), star-forming galaxies that satisfy the high-redshift *UVJ* criterion[45] and with $z_{L68} < z_{cen} < z_{H68}$, where $z_{cen}$ is the redshift of the central and $z_{L68}$ ($z_{H68}$) the lower (upper) 68% confidence limit to the photometric redshift of the putative satellites. Subtracting the background distribution[46] yields an excess of 1234 star-forming satellites, with an additional $\sim 400$ quiescent ones, corresponding to an average of $1 - 2$ satellites per central and a stellar mass distribution consistent with field *BzK*-selected galaxies. As shown in Supplementary Figure 4, the resulting overdensity of star-forming satellite galaxies (hereafter, "satellite halo") becomes clear when averaging over the whole sample and reaches the level of the background at $\gtrsim 30"$ from the central. We estimate the SFR of each satellite thus selected by modelling its rest-frame UV SED with constant star formation stellar population models[53], with and without dust extinction[54]. The obscured SFR of satellites is then given by the difference between the dust-corrected and uncorrected UV SFRs. We then divide the field around the centrals in annuli of width $\Delta \log r = 0.1"$, starting at $r = 1"$ from the central (i.e., twice the seeing and more than 6 times the galaxies' effective radius; see Supplementary Section 4), and sum in each annulus the obscured SFR of satellites to estimate the radial SFR surface density profile of the satellite halo. We model this profile using a combination of a $\beta$-function and a multiplicative, two-parameter "quenching" term of the form $\min(r/r_q, 1)^{\alpha_q}$ (Supplementary Figure 4, B), where r is the distance to the central and $r_q$ and $\alpha_q$ are free parameters, and extrapolate it to the central 2". As the formal error on these parameters contributes negligibly to the uncertainty budget, we fix them to their best-fit values. As the obscured SFR is linearly proportional to the infrared luminosity, the SFR surface density profile is then used as a model of the FIR emission of satellites. Quiescent satellites are not included in this estimate since their average number and stellar mass densities are low, and amount to $(6 \pm 0.7) \times 10^9$ $M_\odot$ when integrated up to the virial radius. If the mass fraction and temperature of the dust in quiescent satellites are the same than for the quiescent centrals (see below), their contribution to the FIR emission is at most 5%, i.e., somewhat smaller than the uncertainties. This value could be even lower if, on the other hand, the quenching of SF in these satellites was mediated by mechanisms that removed the interstellar medium (ISM) at a greater rate than for the centrals.

When the FWHM of the FIR instrumental beam is smaller than the extension of the satellite halo (~20"; this applies to the MIPS, PACS, 250 $\mu$m SPIRE, SCUBA-2, AzTEC, and VLA data), the FIR flux of the central in each band is estimated from the decomposition of the observed signal into a point-source and a 2D extended component based on the SFR profile of the satellite halo, convolved with the instrumental beam. On the other hand, the beam size of SPIRE at 350 and 500 $\mu$m is larger than the extent of the satellite halo, precluding a decomposition of the stacked cutout image into two components. In this case, we only fit the central part of the cutout (within one beam FWHM) with a single point-source and correct the estimate for the fraction of flux in the beam arising from the satellite halo assuming a main-sequence FIR SED[13] for the latter. We note that, in the bands where the two-component decomposition is possible, the two methods yield consistent flux estimates. In both cases the flux of the central galaxy is essentially measured through a point source fit to the unconvolved image. We then take the variance of bootstrap iterations as uncertainty on the flux. Finally, the measured fluxes are corrected for clustering of randomly distributed sources, based on simulations performed on the FIR COSMOS data[55]. The clustering-corrected fluxes and satellite corrections are given in Supplementary Table 1. These, and the total infrared luminosity $L_{IR}$ we find, are a factor ~ 5 lower than those inferred recently for quiescent galaxies at similar redshift and stellar mass[56], but without probing the Rayleigh-Jeans tail of the cold dust emission. Likewise the recovered 1.1 mm is a factor ~4 lower than the contribution of quiescent galaxies to the cosmic infrared background[57]. However, both analyses use a different selection criterion, which, along with our strict treatment of satellite emission, could account for this difference.

The recovered MIR and FIR SED, including the $<3\sigma$ (non-) detections at 100 and 160 $\mu$m, is then modelled with a set of templates of dust grain emission[12] (DL07), assuming $q_{PAH} = 3.19\%$ and $U_{max} = 10^6$ (respectively, the fraction of dust mass in polycyclic aromatic hydrocarbon grains and maximum starlight intensity seen by the dust, relative to the local interstellar radiation field) but with varying $\gamma$ and $U_{min}$ (respectively, fraction of dust enclosed in photodissociation regions and minimum starlight intensity), to estimate the total infrared luminosity $L_{IR}$ and dust mass $M_{dust}$. On the other hand, the dust temperature $T_{dust}$ is determined using a simple modified blackbody model (MBB) with a varying effective emissivity $\beta$, which yields $\beta = 1.8 \pm 0.3$. The derived parameter estimates are given in Supplementary Table 2. We note that $M_{dust}$ estimates could vary substantially between a simple single-temperature MBB fit to the data and that of that of a full dust model like the one provided by DL07. However, when the adopted absorption cross section of the dust distribution ($\kappa_{abs}$) used in the MBB model is consistent both in the normalisation and in the spectral index with that of the full dust model (DL07), the two techniques yield comparable results[58,59].

The major source of uncertainty in the derivation of $M_{dust}$ is the appropriate value of $\kappa_{abs}$, which remains unknown. However, the prescription used in the DL07 model provides $M_{dust}$ (and therefore $M_{gas}$) estimates that are consistent with independent gas measurements of CO excitation in star-forming galaxies at various redshifts[13,60]. Furthermore, the DL07 model has been used to derive $M_{gas}$ estimates for the comparison sample of local quiescent galaxies. Thus, our choice of the DL07 model for our analysis, even if the derived absolute value of $M_{dust}$ remains uncertain, facilitates the comparison between various local and high–z samples and avoids any model dependent systematics.

Finally, we also add to the best-fit DL07 model a power-law radio slope with spectral index $\alpha = 0.8$ and a normalisation given by the FIR-radio correlation[61]. The radio fluxes are however not included in the fit and both the data point and model are shown for reference only. We note that there appears to be a significant excess in the median flux at 1.4 GHz and 3 GHz, compared to the model, corresponding to ~ $5 \times 10^{22}$ W/Hz. On the one hand, this is suggestive of the presence in our sample of low-excitation AGNs[62] and might be a sign of widespread, persistent AGN feedback. In this case the duty cycle should be $f_{BH} \gtrsim 0.5$, since the excess is detected in a median stack. A more accurate value can be estimated under the simple assumption that the AGN activity is randomly distributed, over the time interval ($t_{z=1.4} - t_{z=2.5}$) probed by our sample, in single bursts of varying duration $\Delta t_b$ and constant luminosity set by the low-redshift radio luminosity functions[62]. For each $\Delta t_b$ the radio luminosity is estimated through Monte Carlo simulations by sampling, for each galaxy, the single burst model at its observed redshift. Varying $\Delta t_b$ to match the observed excess luminosity then yields $f_{BH} = \Delta t_b/(t_{z=1.4} - t_{z=2.5}) \sim 0.66$. This would imply that the radio AGN is almost always "on" in these galaxies, as opposed to the local Universe where $f_{BH} < 0.01$[63]. The radio spectral index is ~ −0.82, which is very close to the typical value for SF emission but can also be produced by AGN-related processes[64,65]. On the other hand, the normalisation of the FIR-radio correlation (FRC) has a dispersion of a factor ~ 2[66], potentially lowering the significance of this excess, and in local ETGs the 1.4 GHz flux appears to be weakly correlated with FIR luminosity[67]. However, for an entire galaxy population to deviate from the median relation would imply that the slope and normalisation of the FRC vary substantially with, e.g., star formation efficiency (SFE) or specific SFR (sSFR), which is not substantiated by recent high-redshift studies[68,69,61].

## 3.1 Converting $M_{dust}$ into $M_{gas}$

While $M_{dust}$ can be obtained directly from the FIR emission, the molecular gas mass $M_{gas}$ is a more important quantity in the context of galaxy evolution. Since molecular hydrogen is essentially invisible, $M_{gas}$ is usually extrapolated from other observables, either from sub-mm molecular (e.g., CO) line emission or $M_{dust}$, using conversion factors that are somewhat uncertain. In the case of $M_{dust}$, a gas-to-dust ratio (G/D) is used, which is related to the gas-phase metallicity and to processes of dust creation and destruction. We use here G/D ∼ 95, based on the parameterisation of the G/D as a function of metallicity from Magdis et al. 2012[13] and the assumption of solar metallicity for the gas in the ETGs, which is supported by two lines of indirect evidence: first, the average gas metallicity of star-forming galaxies with the same sSFR is close to (or slightly above) the solar value, up to z ∼ 1.7[70,71]; second, a spectroscopic, rest-frame UV measurement of the metallicity of the last batch of stars in a subset of our ETG sample also yields ∼ $Z_\odot$[27], the stellar metallicity being expected to be slightly lower, but otherwise comparable, to the gas one[72].

The picture is however complicated by processes which can produce dust, such as supernovae, asymptotic giant-branch stars, or accretion of gas-phase metals in the ISM, and those that can destroy it (e.g., X-ray sputtering[73]). While the bulk of cold dust in local ETGs does not appear to originate from stellar evolution[23,74], the G/D of ETGs might nonetheless not follow the same relations as MS galaxies. On the other hand, in local ETGs which have detected $H_2$ gas (i.e., the gas-rich population) the median G/D is somewhat higher (∼ 400), while in the undetected ones the median upper-limit on the G/D is slightly lower (∼ 80) than our adopted value[1]. Overall, the global G/D of local ETGs appears to follow a similar relation to that of star-forming galaxies[75,23,1].

Finally, although knowing the exact state of the gas in our distant ETGs is not necessary for the purposes of this work, we make the simplification that all the hydrogen gas is in molecular ($H_2$) form, rather than atomic (HI) or a mix of both. We base this assumption on the high average ⟨$H_2$/HI⟩ ∼ 2 − 5 ratio seen in local ETGs[76,24,25,77]. Moreover, the $H_2$/HI ratio is expected to increase with redshift[78,79] and gas column density[80]. This ratio appears nevertheless to vary substantially in the local population[81], possibly as a result of differential evolution, with lenticular (S0) galaxies containing higher amounts of $H_2$[82]. We note that, while high-redshift quiescent galaxies tend to have a more disky morphology (i.e., closer to S0s) than local ETGs[83,84,85] (see also Supplementary Section 4), the cold gas in at least some local S0s might have a different origin than that of high-redshift ETGs[86,87].

## 4 Morphologies

There exist several publicly available high-resolution imaging datasets on the COSMOS field, taken with the *Hubble Space Telescope* (HST). The most extensive was obtained with the *Advanced Camera for Surveys* and covers 1.7 degrees at 814 nm[88], corresponding to rest-frame UV wavelengths at z > 1.4. This implies that *HST*/ACS imaging might, for this sample, trace the younger stars produced by the last significant SF episode more than the total stellar mass distribution. Moreover, the existence of colour gradients in high-redshift ETGs is well attested[89,90,91], which implies that their morphological parameters depend on the wavelength at which they are measured. Here we therefore use the much smaller CANDELS survey[92], which benefits from NIR (i.e., rest-frame optical) data at 1.2 and 1.6 $\mu$m. We select the 46 ETGs in our sample whose position overlaps with the *HST*/WFC3 mosaics of the CANDELS survey (a random selection of which is shown in Supplementary Figure 5), as well as 12 medium luminosity stars (14 < V < 16) in the same region. We extract cutouts around the targets from both the F125W and F160W images and combine them using the same method as in Supplementary Section 3, to generate median NIR images of the galaxies and point sources. These are then used in conjunction with the `galfit` code[93] and fitted with a Sérsic profile. This yields a Sérsic index of n = 3.5±0.1 and effective radius of $r_e$ = 1.7±0.1 kpc, consistent with spheroidal morphology and on the stellar mass-size relation for early-type galaxies at this redshift[94,95]. The results of this morphological modelling are shown in Supplementary Figure 6. Furthermore, 40 of these galaxies have individually measured morphologies in the CANDELS catalogue[95], with median values of n ∼ 3 and $r_e$ ∼ 1.3 kpc, respectively. Similar results and high Sérsic indices are also found for the full sample; more details will be presented in a future publication. On the other hand, we note that the in 3 GHz median image the ETG stack appears less concentrated, with $r_e$ ∼ 3.4 kpc for a fit with a fixed n ∼ 1 (exponential) profile. The signal-to-noise ratio of the stacked image is however not sufficient for a two-parameter fit. While n ∼ 1 yields a good fit, it is not clear whether this reflects the structure of the galaxies (or parts thereof) or stems from, e.g., the averaging over the whole sample of randomly oriented jets.

## 5 Hydrodynamical simulations

The cosmological simulation used in this work (Figure 4) is performed in a zoom box focusing on a massive galaxy at the centre of its dark matter halo. The size of the box corresponds to twice the virial radius of the dark halo, without boundary conditions for mass infall. The latter is based on a large-scale cosmological simulation. The simulation of a forming ETG analysed here is from Martig et al. 2009[8], with a complete sample presented in Martig et al. 2012[29]. The evolution of the morphology of both its gas and star components is shown in Supplementary Figure 7. The galaxy starts as disk dominated at high redshift (z > 1.5), then undergoes a series of mergers with mass ratios ranging from 10:1 to 3:1. These mergers have two effects: they convert the stellar structure into a spheroid and compress the gas, increasing the SF and gas consumption rates. Over the 3 Gyr long track used in Figure 4, the stellar morphology is consistently spheroidal, although sometimes disturbed by mergers in their early stages. Over this period, the stellar mass increases from $\sim 6 \times 10^{10}$ $M_\odot$ to $1.2 \times 10^{11}$ $M_\odot$, consistent with our observed sample. About 40% of this increase comes from SF internal to the studied galaxies, while the rest is accreted during mergers with other, smaller galaxies.

The idealised simulations are performed with the Adaptive Mesh Refinement code RAMSES[96]. They are similar to those presented in Martig et al. 2013[32], but have double the resolution, reaching 3 pc in the densest gas regions[2]. The gas fraction coverage has been extended with new simulations, fully identical to the previous ones for all other parameters. These idealised simulations do not follow the formation of a given galaxy in a cosmological context. Instead, the stellar and gas mass, as well as their spatial distributions, are arbitrarily chosen to be representative of a certain type of galaxy. The simulations reach a very high resolution in order to accurately describe the fragmentation of gas in the dense clouds where stars form[32]. They also allow for direct comparisons of the SFE between disk-dominated galaxies and ETGs, with all other parameters being identical. The SFE is determined by the gas surface density, principally, and the rotation curve of the galaxy where the gas is located. In our simulated high-redshift galaxies the critical value is 70 – 100 $M_\odot$ pc$^{-2}$ at the half-mass radius of the gas. This then corresponds to $f_{gas} \sim 25\%$ at high redshift, compared to $f_{gas} = 2 - 3\%$ at $z \sim 0$, due to the expected $\sim 3 - 4$-fold increase in the size of gas disks (i.e., a factor 9 – 16 in surface) in high-redshift ETGs compared to the compact ones seen in local ETGs[76,86,95]. The error bars in Figure 4 correspond to the r.m.s. fluctuation of the SFE between various snapshots of a given simulation. The agreement between these idealised simulations and the cosmological simulation confirms that the process highlighted in the cosmological simulation is not redshift-dependent: while the reduction of the SFE in a spheroid-dominated galaxy (once enough gas has been consumed) happens at about redshift $z \sim 1$ in this simulation, it can just as well occur at somewhat higher redshift, as in our observations. Finally, we note that the idealised simulations do not include diffuse gas reservoirs in the galaxies' outskirts, yielding slightly lower gas masses and thus slightly higher values of the SFE than the cosmological simulation.

## 6  Gas and dust return rate

The amount of gas returned to the ISM by a stellar population depends on both its IMF and the star formation history (SFH). It is generally among the parameters estimated by stellar population synthesis models for single age populations[53,97] (SSP). The typical mass-weighted age of *BzK*-selected quiescent galaxies inferred from spectroscopy is $t_{SF} = 2 - 3$ Gyr[98,27]. We assume a delayed, exponentially declining SFH, arbitrarily starting at z = 10, as the simple parametric form that is closest to the probable "true" SFH of these high-redshift ETGs (rising as the galaxy grows on or above the MS, then decreasing during quenching). In this case, the SFH that would reproduce the SFR inferred from the cold dust emission has $t_{SF} \sim 2.24$ Gyr. The amount of gas returned to the ISM over a certain time span can then be computed easily as the convolution of the SFH and mass loss function due to stellar death. For our chosen SFH and IMF, this would amount to $\sim 4.5 \times 10^{10}$ $M_\odot$, or $37^{+2}_{-1}$% of the observed stellar mass at z = 1.76, and, given the median stellar mass of ETGs in this sample, an average return rate of $1.3^{+1.6}_{-1.3}$ $M_\odot$ yr$^{-1}$ over the last 500 Myr. Similarly, assuming dust yields for solar-metallicity asymptotic giant branch stars[99], $\sim 5 \times 10^8$ $M_\odot$ would have been returned to the ISM, corresponding to $\sim 1\%$ of the returned gas mass. We note that single burst or truncated SFHs, also commonly used in spectral and photometric modelling, would yield similar total gas fractions but negligible return rates at z = 1.76.

The typical best-fit SFH we infer for these ETGs[27] would produce a large amount of post-main-sequence stars at $z \sim 1.76$, which could then contribute significantly to the radiation field to which the dust is exposed. Whether the FIR emission we see is reprocessed radiation from massive, young stars (i.e., SF) or evolved stars then depends on the geometric distribution of the dust, which is unknown. We can, however, estimate the amount of FIR emission produced by non-SF sources in a simple case using the class of delayed exponential SFHs mentioned above, however truncated for the last 500 Myr so as to have no ongoing SF. We then use these SFHs to produce composite stellar population models[53] with which we model the UV-optical SEDs of our ETGs. We assume a standard dust extinction law[100] with a fixed steep slope of $\delta = -0.4$, typical of low sSFR galaxies, and a UV bump amplitude linearly dependent on $\delta$[101]. In this case, we find that the median difference between the integrated flux of models before and after applying extinction, i.e., the maximum luminosity that can arise from evolved stellar populations absorbed by the dust, is $L_{abs} = (9 \pm 5) \times 10^9$ $L_\odot$, or less than half of the measured $L_{IR}$.

Finally, assuming a closed-box evolution to z = 0, an additional $\sim 10^{10}$ $M_\odot$ of gas and $\sim 6 \times 10^8$ $M_\odot$ of dust should be produced by stellar evolution, vastly more than currently observed in local ETGs[23,1] and highlighting the importance of dust and gas removal mechanisms[102,73,63,7] in the evolution of the ISM of ETGs.

# 7 X-ray stacking

We use deep X-ray observations of the COSMOS field by the *Chandra* and *XMM-Newton* observatories[103,104] to constrain the total mass of the halos hosting our sample ETGs. This procedure is similar to that described in Béthermin et al. 2014[105] and Gobat et al. 2015[46], and we refer to these articles for additional information. There are 585 sample ETGs in the *Chandra* COSMOS Survey field and located in zones free from emission. After subtracting both background and point sources, we find a residual flux in the 0.5-2 keV band of $(2.3 \pm 0.6) \times 10^{-17}$ erg cm$^{-2}$ s$^{-1}$, corresponding to a luminosity of $L_X = (5.1 \pm 1.3) \times 10^{42}$ erg cm$^{-2}$ s$^{-1}$ and a mass of $M_{200c} = (1.2 \pm 0.2) \times 10^{13}$ M$_\odot$.

# Data Availability

The data that support the plots within this paper and other findings of this study are available from the corresponding author upon reasonable request.

# References (Cont.)

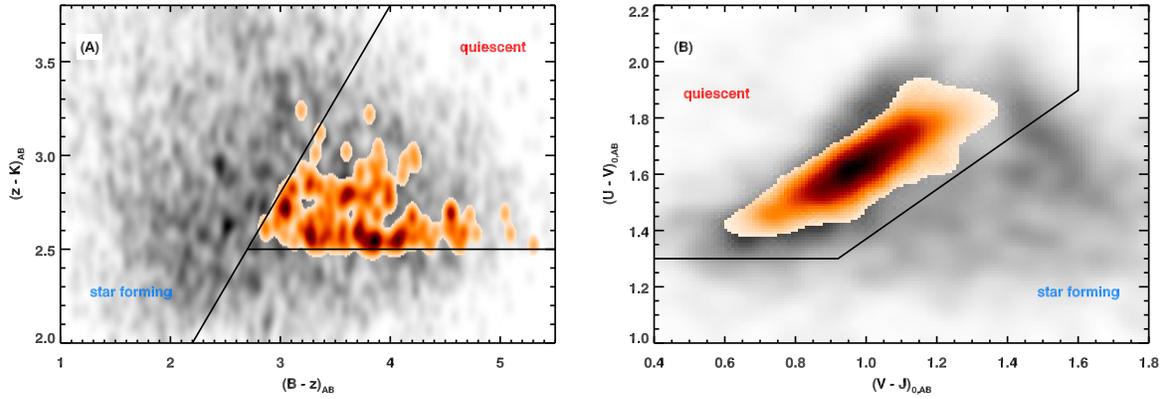

Supplementary Figure 1: Colour-colour diagrams showing the z > 1.4 ETG sample selected using the observed-frame *BzK* (A) and rest-frame *UVJ* (B) criteria. The grey distributions shows the position of sources selected using only the *UVJ* (A) or *BzK* (B) criteria, including, in the case of the latter, sources undetected at 5$\sigma$ in the B or z band.

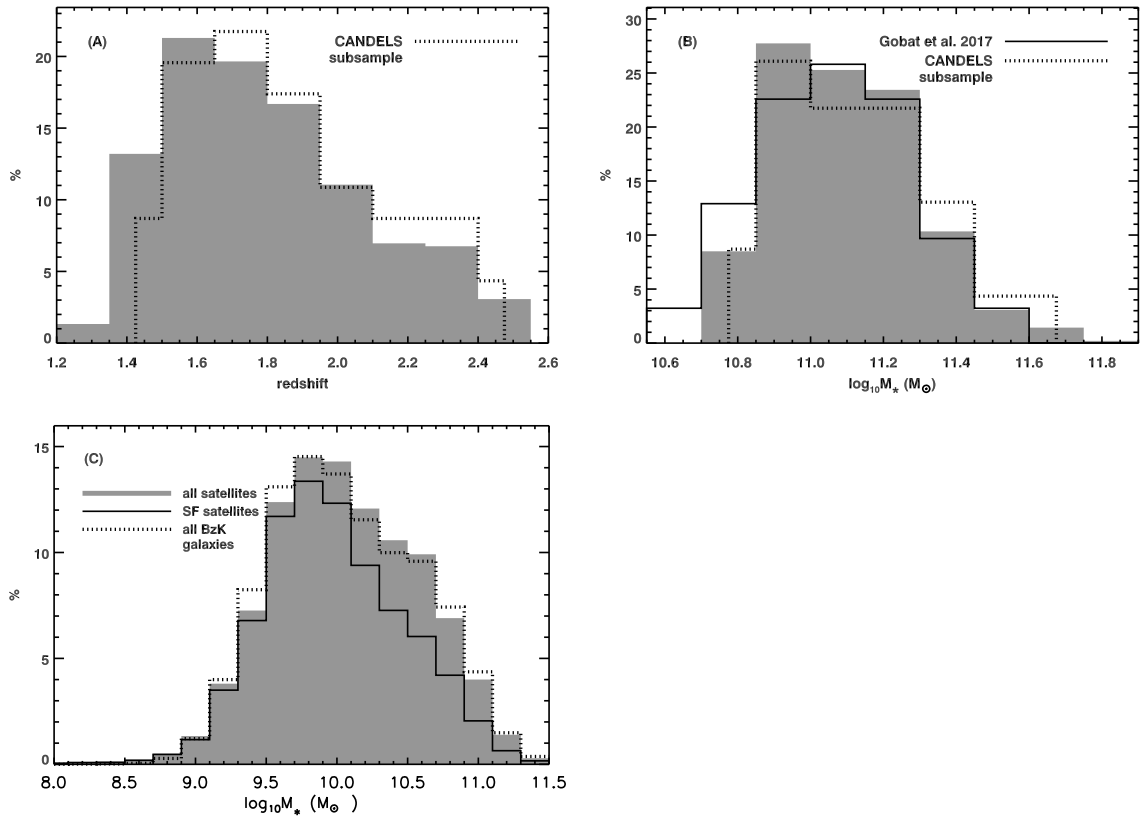

Supplementary Figure 2: Distribution of photometric redshifts (A) and stellar masses of centrals (B) and satellites (C) derived from UltraVISTA photometry. In (A) and (B), the solid histogram shows the distribution of the spectroscopic subsample[27], while the dotted one represents the distribution of sample ETGs in the CANDELS sub-field of COSMOS (see Supplementary Section 4). In (C), the solid histogram shows the distribution of UVJ-selected star-forming satellites, while the dotted one shows, for comparison, the distribution of all BzK-selected galaxies in the COSMOS field, rescaled to match the number of satellites.

Supplementary Table 1: Corrected MIR, FIR, and radio fluxes of the z ~ 1.76 ETGs derived from median stacking, in $\mu$Jy *(top)*, and fraction of flux in the beam emitted by satellites, in % *(bottom)*.

| 24 $\mu$m | 100 $\mu$m | 160 $\mu$m | 250 $\mu$m | 350 $\mu$m | 500 $\mu$m | 850 $\mu$m | 1.1 mm | 10 cm | 22 cm |
|---|---|---|---|---|---|---|---|---|---|
| 6.9 ± 0.6 | 60 ± 25 | 53 ± 22 | 268 ± 36 | 339 ± 39 | 296 ± 26 | 81 ± 23 | 57 ± 16 | 2.0 ± 0.4 | 3.7 ± 0.8 |
| 30 | 40 | 50 | 55 | 60 | 65 | 10 | 5 | < 20 | < 20 |

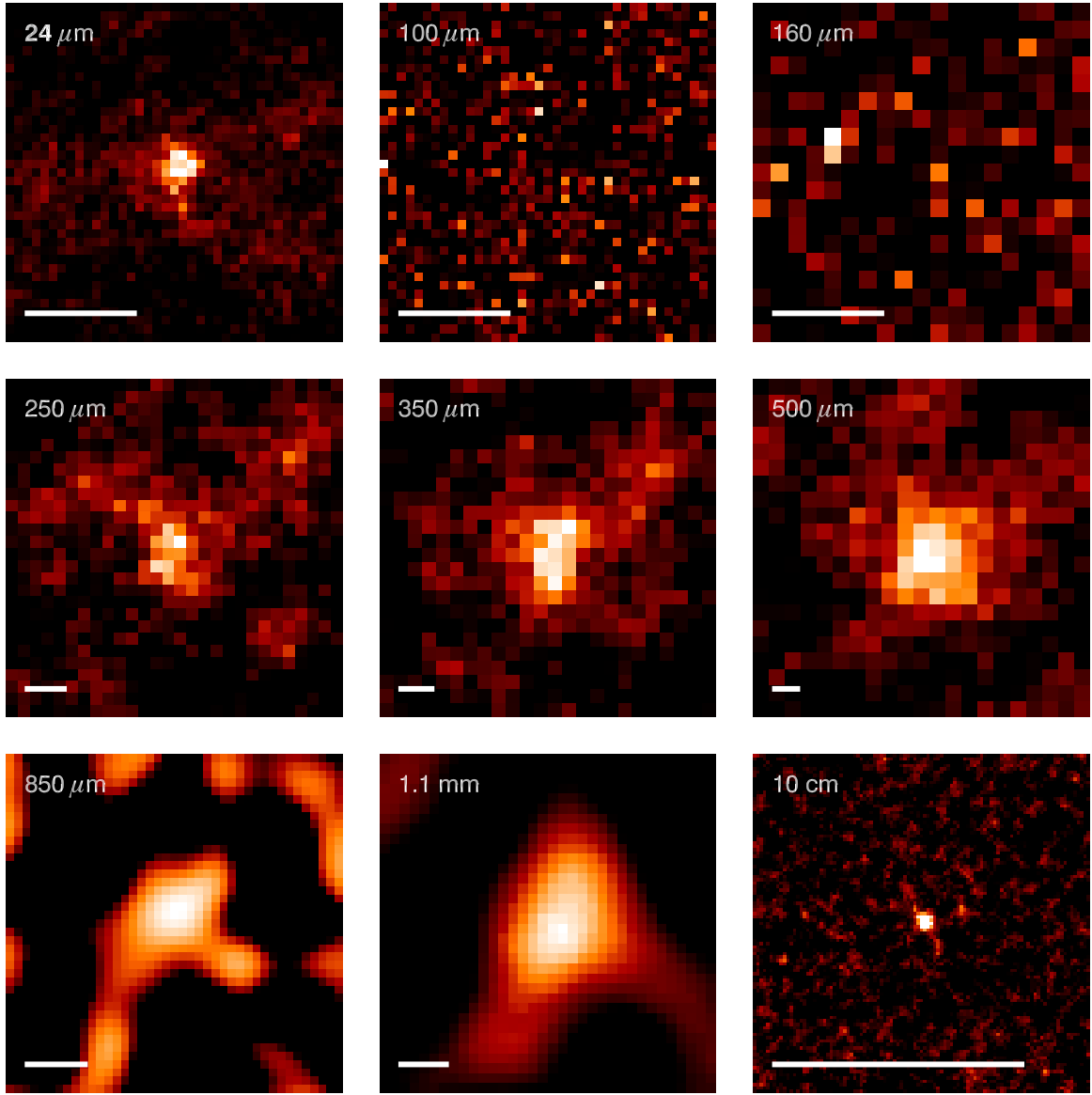

Supplementary Figure 3: Median stacked cutouts at the position of the pBzK+UVJp galaxies, at 24 $\mu$m (*Spitzer/MIPS*), 100 and 160 $\mu$m (*Herschel*/PACS), 250, 350 and 500 $\mu$m *Herschel*/SPIRE, 850 $\mu$m (JCMT/SCUBA-2), 1.1 mm (ASTE/AzTEC) and 3 GHz (JVLA). All cutouts are shown with the same relative flux scale and the white bar in each panel has a length of 15", corresponding to ~130 kpc at z = 1.76.

Supplementary Table 2: Far-infrared properties derived from the stacked FIR emission of the z ~ 1.76

| log $L_{IR}$ [a] | SFR[b] | $\langle U \rangle$[c] | $T_{dust}$[d] | log $M_{dust}$[e] | log $M_{gas}$[f] | log $M_\star$[g] | $f_{gas}$[h] |
|---|---|---|---|---|---|---|---|
| $L_\odot$ | $M_\odot$ yr$^{-1}$ | | K | $M_\odot$ | $M_\odot$ | $M_\odot$ | |
| 10.46 ± 0.14 | $4.8^{+1.8}_{-1.3}$ | 2.2 ± 1.1 | $23.8^{+1.6}_{-1.8}$ | 8.00 ± 0.2 | 9.98 ± 0.2 | 11.08 ± 0.09 | $7.4^{+4.6}_{-3.0}$% |

[a] Total IR luminosity integrated from 8 to 1000 $\mu$m. [b] Star formation rate. [c] Mean starlight intensity. [d] Dust temperature. [e] Dust mass. [f] Hydrogen gas mass (see Methods). [g] Stellar mass. [h] Gas fraction, $M_{gas}/(M_\star + M_{gas})$

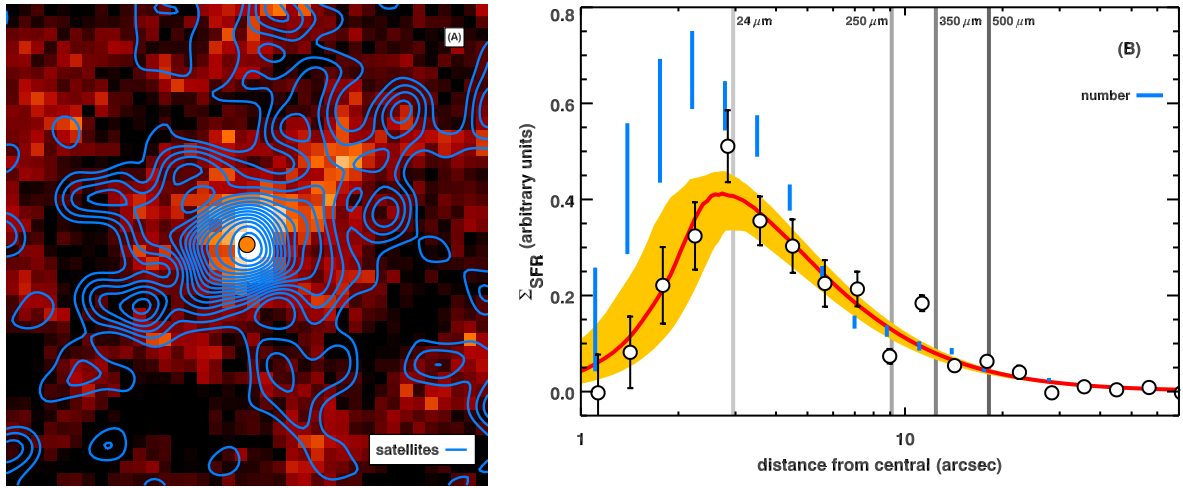

Supplementary Figure 4: *A):* median number density of photometrically selected satellites (blue contours), compared to the stacked 350 $\mu$m *Herschel*/SPIRE stack (red map) and with respect to the position of the ETG centrals (orange dot). The cutout shows a 60'×60' field. *B):* Average SFR surface density ($\Sigma_{SFR}$) profile of the satellites of $z \sim 1.76$ ETGs (circles) with, in blue, the number surface density rescaled to match $\Sigma_{SFR}$ at large radii, for convenience. In both cases the error bars correspond to the median absolute deviation of the surface density in each bin. The red line and yellow shaded region show the best-fit "quenched" $\beta$-model to $\Sigma_{SFR}$, with uncertainties. The grey vertical lines show the half-width at half-maximum of the instruments' beams from 24 $\mu$m to 500 $\mu$m.

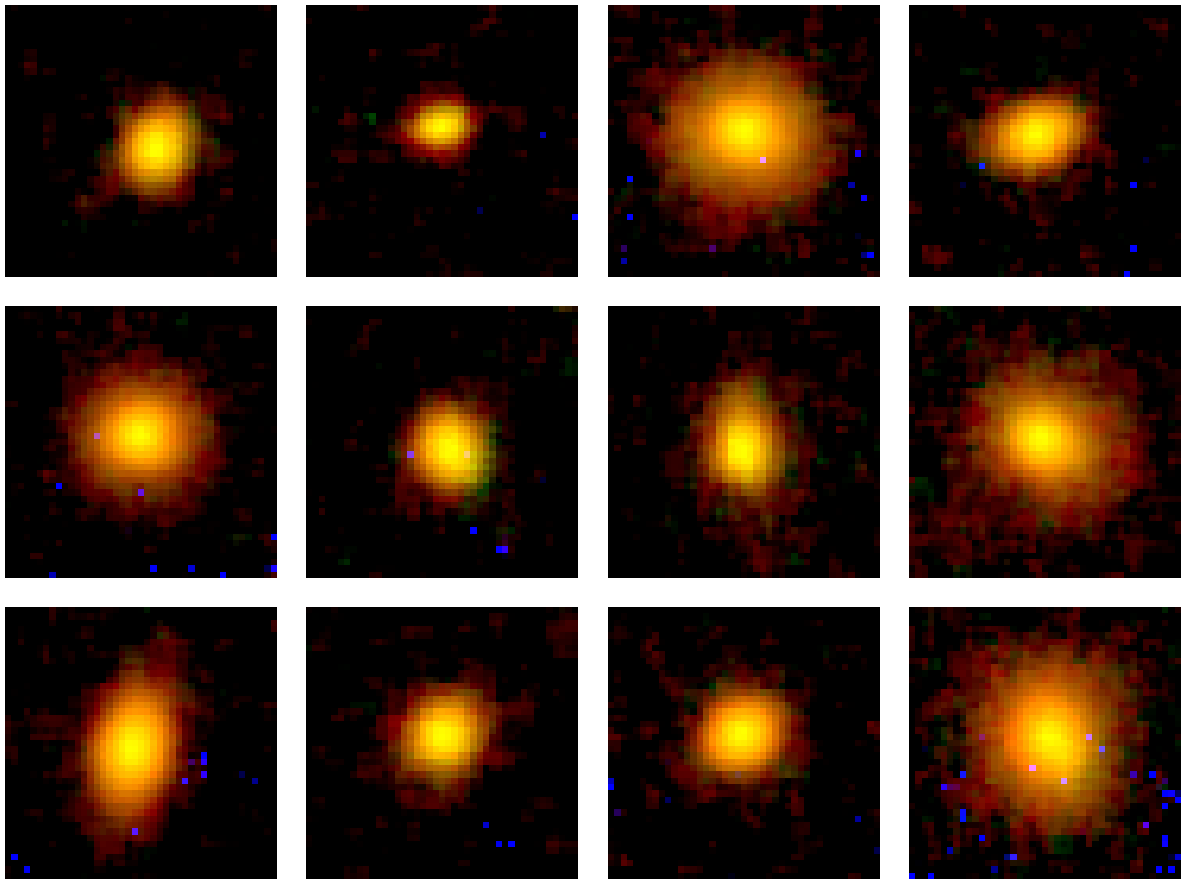

Supplementary Figure 5: RGB-composite (*HST*/WFC3 F160W, F125W, F606W) cutout images of sample ETGs in the CANDELS[92] field. The F125W and F606W images have been smoothed with a Gaussian kernel to match the resolution of the F160W data. Each cutout is 2.4" (or ~20 kpc) on the side.

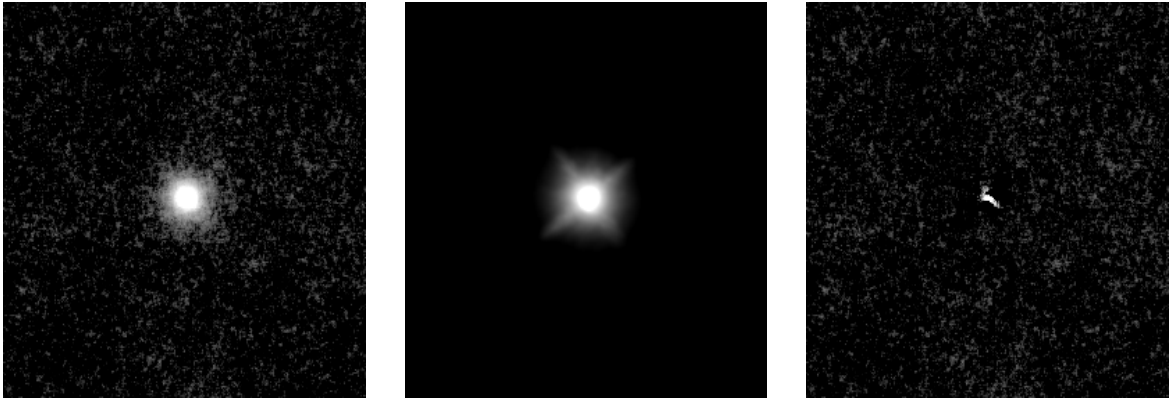

Supplementary Figure 6: Median *HST*/WFC3 F160W stack of the ETGs in the CANDELS field (left), with best-fit Sérsic model (centre) and residuals after subtraction (right), for the same scale and limits.

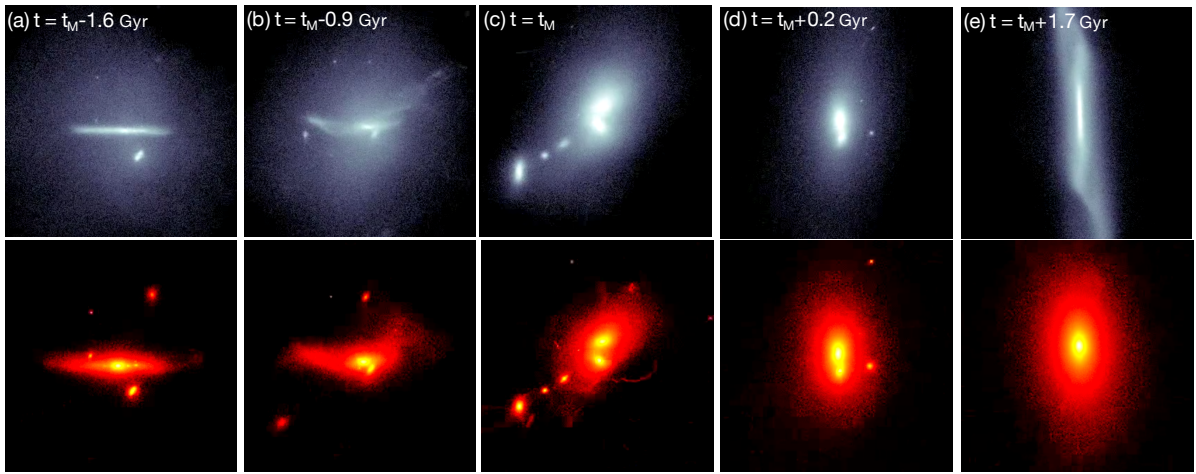

Supplementary Figure 7: The cosmological simulation analysed here[8] models the formation and evolution of an ETG. Starting disk-dominated at redshifts $z \sim 2$ and above (a), this galaxy experiences several mergers that disrupt its stellar disk (b) and convert it into a spheroid, with the last significant mergers shortly before z=1 (c). The mergers compress the gas toward the centre (d), inducing bursts of star formation and gas consumption and reducing the gas fraction to about 20%. The gas then rapidly settles back in a thin rotating disk, but the lack of gravitational coupling to a co-rotating stellar disk quenches star formation with a SFR 5-10 times lower than in a spiral galaxy with the same gas content (e). The captions show the gas (top) and star (bottom) surface density at different redshift, over sizes of $20 \times 20$ kpc.

16